\documentclass[aip,apl,twocolumn,superscriptaddress]{revtex4}
\usepackage{amssymb}
\usepackage{amsmath}
\usepackage{graphicx}
\usepackage{natbib}

\begin{document}
\textheight = 63\baselineskip
\title{Single charge sensing and transport in double quantum dots fabricated from commercially grown Si/SiGe heterostructures}

\author{C. Payette}
\affiliation{Department of Physics, Princeton University, Princeton, NJ 08544, USA}
\author{K. Wang}
\affiliation{Department of Physics, Princeton University, Princeton, NJ 08544, USA}
\author{P. J. Koppinen}
\affiliation{Department of Physics, Princeton University, Princeton, NJ 08544, USA}
\author{Y. Dovzhenko}
\affiliation{Department of Physics, Princeton University, Princeton, NJ 08544, USA}
\author{J. C. Sturm}
\affiliation{Department of Electrical Engineering, Princeton University, Princeton, NJ 08544, USA}
\affiliation{Princeton Institute for the Science and Technology of Materials (PRISM), Princeton University, Princeton,
New Jersey 08544, USA}
\author{J. R. Petta}
\affiliation{Department of Physics, Princeton University, Princeton, NJ 08544, USA}
\affiliation{Princeton Institute for the Science and Technology of Materials (PRISM), Princeton University, Princeton,
New Jersey 08544, USA}

\date{\today}

\begin{abstract}
We perform quantum Hall measurements on three types of commercially available modulation-doped Si/SiGe heterostructures to determine their suitability for depletion gate defined quantum dot devices. By adjusting the growth parameters, we are able to achieve electron gases with charge densities 1--3$\times$10$^{11}/$cm$^{2}$ and mobilities in excess of 100,000 cm$^{2}$/Vs. Double quantum dot devices fabricated on these heterostructures show clear evidence of single charge transitions as measured in dc transport and charge sensing and exhibit electron temperatures of 100 mK in the single quantum dot regime.

\end{abstract}


\maketitle
Possible applications in quantum information processing have motivated current research into silicon-based single electron devices \cite{morello10,borselli11,simmons11}. Due to the relatively small spin-orbit coupling strength and dilute concentration of spin-carrying nuclear isotopes, the coherence times of electron spins in Si quantum dots are expected to be greatly extended relative to GaAs \cite{petta05,hanson07}. Progress in developing depletion mode Si/SiGe quantum dot devices has however been limited by the availability of high quality Si/SiGe two-dimensional electron gases (2DEGs) with low charge density and high carrier mobility. Challenges include gate leakage, charge instabilities, switching noise and a high density of threading defects stemming from the growth of the relaxed buffer substrate \cite{abstreiter85,fitzgerald92,ismail95,schaffler97,wild10}.

In this letter, we systematically explore the relationship between the heterostructure growth profile and 2DEG quality by varying 2DEG depths and doping levels. We identify several heterostructure growth profiles where the 2DEG has low electron density, $n$, high electron mobility, $\mu$, and shows no evidence of parallel conduction attributable to charge accumulation near the Si cap layer. Double quantum dots fabricated on the most promising wafers are investigated using dc transport and quantum point contact based charge sensing. When tuned to the single dot regime, we observe clear signatures of single electron charging and low electron temperatures, $T_e$ = 100 mK.  In the double dot regime, we clearly resolve single dot and interdot transitions in charge sensing, which suggests these materials may be useful platforms for developing spin qubits in Si.

\begin{figure}[t!]
\begin{center}
		\includegraphics[scale=0.95]{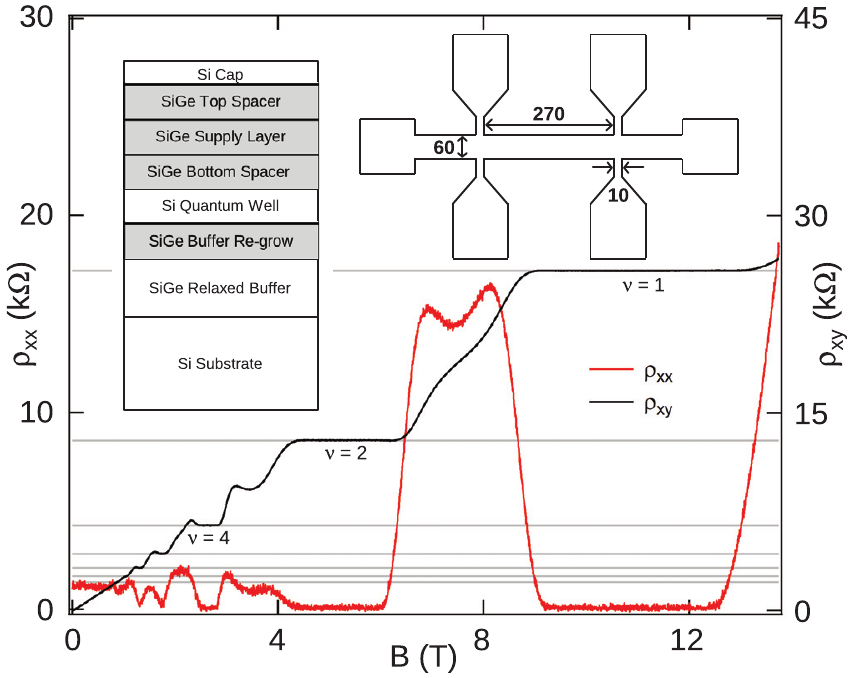}
\caption{\label{fig1} (Color online)
Typical Hall data set from which we extract $\mu$ = 96,000 cm$^{2}$/Vs and $n$ = 2.5 $\times$ $10^{11}$/cm$^{2}$. Wafer details are given in main text. Horizontal lines identify the visible quantum Hall plateaus. Upper left inset: Heterostructure growth profile. Upper right inset: Hall bar geometry with dimensions given in microns.}
\end{center}	
\end{figure}

We measure the transport properties of Si/SiGe heterostructures grown using chemical vapor deposition by Lawrence Semiconductor Research Laboratories. Three main heterostructure growth profiles are investigated, based on previous reports in the literature \cite{hayes09, thalakulam10, yao09thesis}. Layer thicknesses and doping profiles are listed in Table \ref{table1}. Relaxed buffers are grown on Si substrates by varying the Ge content from 0 to 30\% over a thickness of 3 $\mu$m. A 300 nm thick layer of Si$_{0.7}$Ge$_{0.3}$ is grown on the virtual substrate before it is polished. After polishing, the wafers are completed by growing a 225 nm thick Si$_{0.7}$Ge$_{0.3}$ layer, followed by a strained-Si quantum well, a Si$_{0.7}$Ge$_{0.3}$ bottom spacer, a phosphorus-doped Si$_{0.7}$Ge$_{0.3}$ supply layer, a Si$_{0.7}$Ge$_{0.3}$ top spacer, and a Si cap. The growth structure is shown in the upper left inset of Fig.\ \ref{fig1}.

\begin{table} 
\begin{ruledtabular}
\begin{tabular}{|cccc|}
\hline
Layer & Series 1 & Series 2 & Series 3 \\
\hline
Si Cap (nm) & 7.5 & 9 & 11 \\
\hline
SiGe Top Spacer (nm) & 25 & 45 & 25\\
\hline
SiGe Supply Layer (nm)	     & 20 & 2.6 & 2.5\\
Doping Range (/cm$^{3}$)	& 2--10x10$^{17}$ & 6--50x10$^{17}$ & 5--50x10$^{17}$\\
\hline
SiGe Bottom Spacer (nm)	     & 5 or 10 & 22  & 22\\
\hline
Si Quantum Well (nm)	     & 15 & 18 & 10\\
\hline
SiGe Buffer Re-grow (nm) & 225 & 225  & 225\\
\hline
SiGe Relaxed Buffer ($\mu$m) & 3 & 3 & 3\\
\hline

\end{tabular}
\end{ruledtabular}
\caption{Layer thicknesses for the three different heterostructure growth profiles studied in this report.
        \label{table1}}
\end{table}

We perform magnetotransport measurements on Hall bars fabricated from the wafers (see upper right inset of Fig.\ \ref{fig1}). Ohmic contacts are made by thermally evaporating a 20/1/30/1/70 nm stack of Au/Sb/Au/Sb/Au and annealing at 390 $^{\circ}$C for 10 min. Low frequency ac lock-in techniques are used to simultaneously measure the longitudinal voltage, $V_{xx}$, and Hall voltage, $V_{xy}$, as a function of field, $B$, with a 10 nA current excitation. Charge density is extracted from the low field Hall response ($B$ $<$ 1 T) and the mobility is extracted from measurements of the zero field $V_{xx}$. All measurements were performed in a top-loading dilution refrigerator equipped with a 14 T superconducting magnet. The base temperature of the cryostat is 35 mK.

Figure \ref{fig1} shows typical Hall data measured on a series 1 sample after illumination with a red light-emitting diode (LED) for 1 min. The wafer has a 10 nm bottom spacer layer and doping level of 6 $\times$ $10^{17}/$cm$^{3}$. Clear integer quantum Hall plateaus are visible for filling factors, $\nu$ = 1, 2, 4, 6, 8, 10 and 12 at values of $\rho_{xy} = h/(\nu e^{2})$. As is the case in Fig.\ \ref{fig1}, we always observe more Hall plateaus for even values of $\nu$ than for odd values. This is consistent with previous reports that the valley splitting is smaller than the Zeeman splitting in similar Si/SiGe heterostructures in the quantum Hall regime \cite{sailer10}. The data in Fig.\ \ref{fig1} also reveal vanishing minima in $\rho_{xx}$ at fields corresponding to the $\nu$ = 1, 2, and 4 plateaus. The vanishing quantum Hall minima in $\rho_{xx}$, coupled with the high mobilities observed, confirm that the 2DEG is the only conduction channel, which is an important prerequisite for operating quantum dot devices. Overshoots in $\rho_{xy}$ on the low field side of the resolvable quantum Hall plateaus for $\nu > 2$ have been attributed to the co-existence of incompressible strips with different filling factors \cite{sailer10}.

\begin{figure} [tr!]
\begin{center}
		\includegraphics[scale = 0.95]{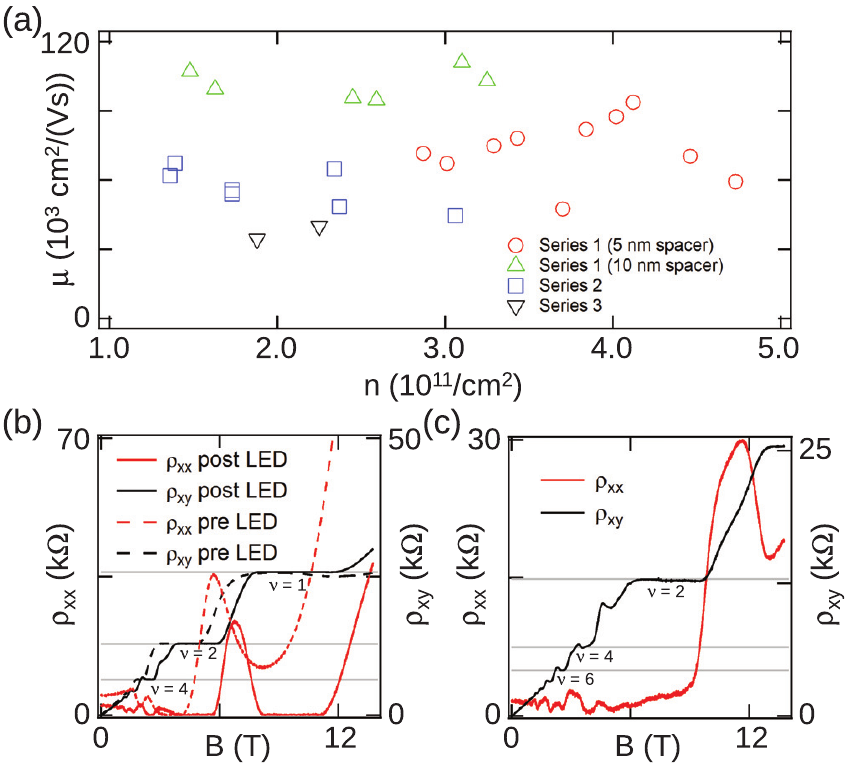}
\caption{\label{fig2} (Color online) (a) Carrier mobility plotted as a function of charge density for each wafer series. (b) Hall data sets recorded before and after illumination with a LED for a series 2 wafer with a doping level of 3 $\times$ 10$^{18}/$cm$^{3}$, resulting in $\mu$ = 30,000 cm$^{2}$/Vs and $n$ = 1.8 $\times$ 10$^{11}/$cm$^{2}$ before illumination and $\mu$ = 49,000 cm$^{2}$/Vs and $n$ = 2.4 $\times$ 10$^{11}/$cm$^{2}$ after illumination. (c) A series 1 wafer (5 nm spacer) with a doping level of 6 $\times$ 10$^{17}/$cm$^{3}$ does not show clear zeros in $\rho_{xx}$ after illumination with a LED. $\mu$ = 48,000 cm$^{2}$/Vs and $n$ = 3.7 $\times$ 10$^{11}/$cm$^{2}$ for this sample.}
\end{center}	
\end{figure}

Hall data were recorded from 16 wafers after illumination with a LED, resulting in the scatter plot of mobility as a function of density shown in Fig.\ \ref{fig2}(a). Hall bars fabricated on the same chip showed variations in carrier density of less than 5\% and carrier mobility of less than 10\%, provided that there was no evidence of parallel conduction. The data in Fig.\ \ref{fig2}(a) indicate that the specific heterostructure growth profile can have a large impact on 2DEG quality. Series 3 wafers support low density 2DEGs with $n$ $\sim$ 2 $\times10^{11}$/cm$^{2}$, but the mobility is comparably poor, $\mu$ $<$ 40,000 cm$^{2}$/Vs. While series 2 wafers support low density 2DEGs with $n$ = 1--3 $\times$ $10^{11}\mathrm{/cm}^{2}$, the mobilities are moderate, $\mu$ $\sim$ 70,000 cm$^{2}$/Vs. Series 1 wafers with a 5 nm thick bottom spacer have comparably high mobilities (typically above 60,000 cm$^{2}$/Vs and reach a maximum of $\sim$ 90,000 cm$^{2}$/Vs). However densities below 3 $\times$ $10^{11}$/cm$^{2}$ were not attainable. Increasing the bottom spacer thickness of the series 1 wafers to 10 nm results in $\mu$ $\sim$ 100,000 cm$^{2}$/Vs with charge densities in the range of 1--3 $\times$ $10^{11}$/cm$^{2}$, indicating that this series is the most promising for the fabrication of quantum dot devices.

For each measured sample, we attempted to perform Hall measurements prior to illumination, but the Ohmic contact resistance for low density samples was on the order of 1 M$\Omega$, independent of wafer series. Illumination typically reduced the contact resistances to a few k$\Omega$ allowing us to make reliable Hall measurements. Furthermore, for samples with low Ohmic contact resistance prior to illumination, Hall measurements revealed non-vanishing quantum Hall minima in $\rho_{xx}$ for all but the highest density samples. As a specific example, it is evident in the  data in Fig.\ \ref{fig2}(b) that near the $\nu = 1$ plateau $\rho_{xx}$ is finite prior to illumination and vanishes after illumination. However, illumination does not guarantee vanishing quantum Hall minima in $\rho_{xx}$. For each series, samples with the lowest densities did not exhibit zeros in the longitudinal resistance after illumination [see Fig.\ \ref{fig2}(c)]. Optimal quantum dot device performance will most likely be achieved by illuminating the samples prior to measurements and by selecting material with a moderate density, rather than the lowest measurable density within a series.

\begin{figure}[t!]
\begin{center}
			\includegraphics[scale = 0.95]{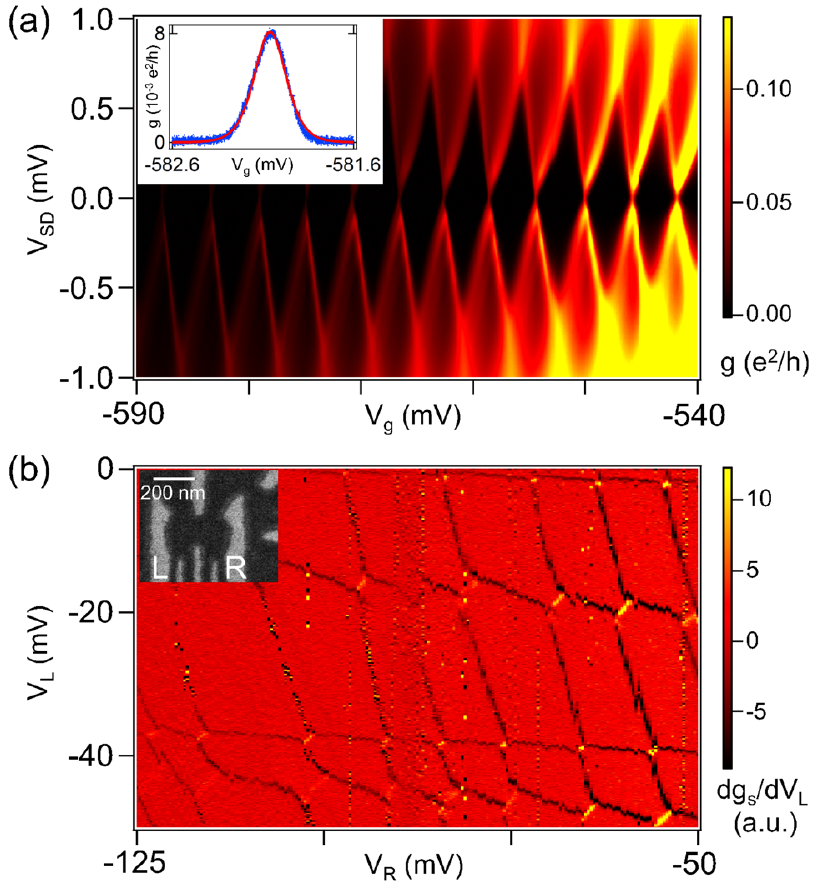}
\caption{\label{fig3} (Color online) (a) Colorscale plot of quantum dot conductance, $g$, as a function of $V_{g}$ and $V_{SD}$. Clear Coulomb diamonds are observed when the device is configured as a single dot. Inset: Conductance as a function of $V_g$ for $V_{SD}$ = 0.  (b) Derivative of the charge sensor conductance, $dg_{S}/dV_{L}$, measured as a function of $V_{L}$ and $V_{R}$. Black transitions (negative slopes) correspond to charge transitions that change the total number of electrons, while yellow transitions (positive slopes) correspond to interdot charge transitions. Inset: Scanning electron micrograph of a double quantum dot similar to the one measured.}
\end{center}	
\vspace{-0.5 cm}
\end{figure}

Based on the quantum Hall characterization, we fabricated quantum dot devices on a series 1 wafer with a 5 nm spacer thickness and a doping level of 8 $\times$ 10$^{17}/$cm$^{3}$, which yields $n$ = 3.4 $\times$ 10$^{11}/$cm$^{2}$  and $\mu$ = 78,000 cm$^{2}$/Vs. Low-leakage Pd top gates are used to define the double quantum dot [see inset of Fig.\ 3(b)]. Figure 3(a) shows the conductance, $g$, through the device measured as a function of gate voltage, $V_g$, and bias voltage, $V_{SD}$, yielding clear Coulomb diamonds in the single dot regime. The absence of abrupt switching during this 14 hour scan demonstrates the excellent stability of the device in this regime. By fitting to the Coulomb blockade peak shown in the inset of Fig. \ \ref{fig3}(a), we extract an electron temperature $T_e$ = 100 mK \cite{beenakker91}. We consistently observe electron temperatures in the range of 100--150 mK, which suggests that previous reports of high electron temperatures in Si quantum dots may be due to sample quality or electrical filtering \cite{thalakulam11}.

We are able to form a double quantum dot by energizing all of the depletion gates on the device. The device is probed using a proximal quantum point contact as a charge sensor. The derivative of the charge sensor conductance, $dg_{s}/dV_{L}$, is plotted in Fig.\ \ref{fig3}(b) as a function of the gate voltages $V_{L}$ and $V_{R}$. We observe clear charging lines that correspond to interdot charge transitions, where the total electron number is fixed, and transitions with the leads, where an electron is either added or removed from the double quantum dot. In contrast to the cleanest GaAs double dot devices \cite{petta05}, the  stability diagram shows transitions with three distinct negative slopes, indicative of accidental triple quantum dot formation \cite{gaudreau06}. Attempts to completely empty the double dot of electrons were unsuccessful due to the onset of switching noise at large negative biases. Further modification of the depletion gate design may allow us to reach the few electron regime at more moderate values of the depletion gate voltage.

In summary, we have studied three heterostructure growth profiles in an effort to identify low density, high mobility samples, ideal for fabricating quantum dot devices. We fabricated a double quantum dot device that shows stable Coulomb diamonds in the single quantum dot regime. When tuned to the double dot regime the device exhibits clear single electron transitions in charge sensing. Improvements in the depletion gate geometry may allow us to reproducibly enter the few-electron regime in charge sensing and transport, as required for quantum manipulation of singlet-triplet spin qubits \cite{petta05}.

CP acknowledges partial support from FQRNT. Research at Princeton was supported by the Sloan and Packard Foundations, the NSF through the Princeton Center for Complex Materials (DMR-0819860) and EMT/QIS award (CCF-0829872), and the United States Department of Defense. The views and conclusions contained in this document are those of the authors and should not be interpreted as representing the official policies, either expressly or implied, of the U.S. Government.


%

\end{document}